\documentclass[prd,notitlepage,showpacs,nofootinbib,superscriptaddress]{revtex4-2}

\usepackage[utf8]{inputenc} 

\usepackage{amsmath}
\usepackage{amssymb}
\usepackage{float}
\usepackage{comment}
\usepackage{slashed}
\usepackage[normalem]{ulem}
\usepackage{bbold}
\usepackage{cancel}
\usepackage{graphicx}
\usepackage{booktabs}
\usepackage{multirow}
\usepackage{rotating}
\usepackage{caption}
\usepackage{subcaption}
\usepackage{tabulary}

\usepackage[usenames,dvipsnames]{color}

\newcommand {\ignore}[1]{}

\usepackage[colorinlistoftodos]{todonotes}
\usepackage[colorlinks=true,linkcolor=darkblue,citecolor=darkblue,urlcolor=darkblue, pdfborder={0 0 0}]{hyperref}
\usepackage[normalem]{ulem}

\usepackage{caption}
\usepackage{subcaption}
\definecolor{linkcolor}{rgb}{0,0,0.8}

\definecolor{darkgreen}{rgb}{0,0.5,0}
\definecolor{darkred}{rgb}{0.6,0,0}
\definecolor{brown}{rgb}{0.59, 0.29, 0.0}
\definecolor{mightnightblue}{RGB}{25,25,112}
\definecolor{darkblue}{rgb}{0,0,0.8}

\newcommand {\darkblue} {\color{darkblue}}

\def\SM{$\mathrm{SU(3)_c} \otimes \mathrm{SU}(2)_L \otimes \mathrm{U}(1)_Y $}

\newcolumntype{K}[1]{>{\centering\arraybackslash}m{#1}}
\def\gsim{\raise0.3ex\hbox{$\;>$\kern-0.75em\raise-1.1ex\hbox{$\sim\;$}}}
\def\lsim{\raise0.3ex\hbox{$\;<$\kern-0.75em\raise-1.1ex\hbox{$\sim\;$}}}

\newcommand{\AddrIEAP}{%
Institute of Experimental and Applied Physics, \\ Czech Technical University in Prague, \\
Prague 160 00, Czech Republic}

\usepackage[force]{feynmp-auto}

\usepackage{colortbl}

\begin{document}

\title{\darkblue One-loop proton decay from Peccei-Quinn symmetry}

\author{\textbf{H.~B. C\^amara}}\email{henrique.camara@cvut.cz}
\affiliation{\AddrIEAP}

%%%%%%%%%%%%%%%%%%%%%%%%%%%%%%%%%%%%%%%%%%%%%%%%%%%%%%%%%%%%%%%%%%%%%%%%%%%%%
\begin{abstract}
\vspace{0.5cm}
\begin{center}
{ \center  \bf ABSTRACT}\\    
\end{center}
We promote the accidental $B+L$ symmetry of the Standard Model to a Peccei-Quinn~(PQ) symmetry while realizing spontaneous proton decay radiatively. The PQ anomaly sector consists of vector-like quarks~(VLQs), providing a Kim-Shifman-Vainshtein-Zakharov-type axion solution to the strong CP problem. After spontaneous PQ breaking, a residual $\mathcal{Z}_2$ symmetry remains, which forbids tree-level proton decay. The VLQs required to generate the QCD anomaly, together with scalar mediators, odd under the residual $\mathcal{Z}_2$, induce one-loop proton decay through the effective operator $u_R u_R d_R e_R$. We study its ultraviolet completions, featuring distinct vector-like fermion and scalar representations, and show that the resulting models lead to distinct predictions for the axion-to-photon coupling, testable in helioscope and haloscope experiments. We investigate proton and neutron decay pheneomenology focusing on the channel $p \rightarrow e^+ \pi^0$. We also discuss axion dark matter in pre- and post-inflationary cosmology.
\end{abstract}
%%%%%%%%%%%%%%%%%%%%%%%%%%%%%%%%%%%%%%%%%%%%%%%%%%%%%%%%%%%%%%%%%%%%%%%%%%%%%

\maketitle
\noindent

%%%%%%%%%%%%%%%%%%%%%%%%%%%%%%%%%%%%%%%%%%%%%%%%%%%%%%%%%%%%%%%%%%%%%%%%%%%%%
\section{Introduction}
\label{sec:intro}
%%%%%%%%%%%%%%%%%%%%%%%%%%%%%%%%%%%%%%%%%%%%%%%%%%%%%%%%%%%%%%%%%%%%%%%%%%%%%

Proton stability is ensured in the Standard Model~(SM) by the accidental baryon number~$B$ and lepton number~$L$ symmetries.
However, in the SM effective field theory, higher-dimensional operators consistent with the SM gauge symmetry generically violate $B$ and $L$.
Already at dimension five, the unique Weinberg operator $(\ell_L \Phi)(\ell_L \Phi)/\Lambda$, involving the lepton and Higgs doublet, where $\Lambda$ is the typical new physics scale, violates lepton number by two units~\cite{Weinberg:1979sa}. After electroweak~(EW) symmetry breaking when $\Phi$ acquires a non-zero vacuum expectation value~(VEV), this operator generates Majorana neutrino masses. At dimension six, the leading baryon-number-violating effects arise from four-fermion operators involving three quark fields and one lepton field $qqq\ell/\Lambda^2$~\cite{Weinberg:1979sa,Wilczek:1979hc,Abbott:1980zj,Kobach:2016ami,Grzadkowski:2010es}. These operators mediate two-body nucleon decay processes that violate $B$ and $L$ each by one unit, implying $B+L$ violation by two units, while conserving $B-L$.

From an ultraviolet~(UV) perspective, such operators arise naturally in Grand Unified Theories~(GUTs)~\cite{Pati:1973uk,Georgi:1974sy,Fritzsch:1974nn}. A prime example is the minimal SU(5) GUT, where the exchange of heavy gauge bosons generates $qqq\ell/\Lambda^2$ at tree level~\cite{Georgi:1974sy}. These operators induce baryon-number-violating nucleon decay modes, most notably the golden channel $p\rightarrow e^+\pi^0$ (for a recent review on $B$ and $L$ violation see Ref.~\cite{Takhistov:2026aln}). Experimentally, Super-Kamiokande~(Super-K), reports the most stringent bound on the partial proton mean lifetime: $\tau(p\rightarrow e^+\pi^0) > 2.4\times10^{34}$ years
at 90$\%$ CL~\cite{Super-Kamiokande:2020wjk}. This constraint pushes the GUT proton decay mediator mass scale to $\Lambda\sim \mathcal{O}(10^{16})$~GeV. Future large-volume detectors, including Hyper-Kamiokande~(Hyper-K)~\cite{Hyper-Kamiokande:2018ofw}, DUNE~\cite{DUNE:2020ypp}, and JUNO~\cite{JUNO:2021vlw}, are expected to substantially extend the sensitivity to nucleon decay and provide significantly improved probes of the $B+L$-violating processes predicted in GUT scenarios.

\newpage

Proton decay may itself be predominantly a radiative phenomenon. Namely, Ref.~\cite{Helo:2019yqp} systematically classifies one-loop decomposition and UV completions of the dimension-six $B+L$-violating operators $qqq\ell/\Lambda^2$. In these scenarios, the mediators responsible for proton decay, such as scalar leptoquarks and/or vector-like quarks~(VLQs), can lie at the TeV scale, opening the possibility of direct collider tests. Additionally, radiative proton-decay frameworks have been studied in connection to neutrino masses and dark matter~(DM). In constructions based on the scotogenic idea~\cite{Tao:1996vb,Ma:2006km}, the same dark-sector states responsible for generating neutrino masses radiatively can also mediate proton decay, linking the smallness of neutrino masses and the longevity of the proton~\cite{Nomura:2024zca,Kang:2024oyf,IBeneito:2025nby,Beneito:2025fzf}. Furthermore, in Ref.~\cite{Kumar:2025aek}, the accidental $B+L$ symmetry of the SM is promoted to a global U(1)$_{B+L}$ symmetry, which is subsequently spontaneously broken to a residual $\mathcal{Z}_4$ symmetry, stabilizing a viable weakly interacting massive particle~(WIMP) DM candidate. Although proton decay is forbidden at tree level, it is generated via $\mathcal{Z}_4$-odd mediators at one-loop.

Another longstanding issue of the SM concerns the strong CP phase $\bar{\theta}$, which encodes CP violation in Quantum Chromodynamics (QCD).
Experimental searches for the neutron electric dipole moment impose the stringent bound $|\bar{\theta}| \lesssim 10^{-10}$~\cite{Pendlebury:2015lrz,Baker:2006ts}, yet the SM offers no theoretical explanation for its smallness, giving rise to the strong CP problem.
A particularly elegant solution relies on the Peccei-Quinn~(PQ) mechanism~\cite{Peccei:1977hh,Peccei:1977ur}, which introduces a global QCD-anomalous U(1)$_{\rm PQ}$ symmetry whose spontaneous breaking yields a pseudo-Goldstone boson, the axion~\cite{Weinberg:1977ma,Wilczek:1977pj}.
Non-perturbative QCD effects generate an axion potential whose minimum dynamically relaxes $\bar{\theta}\rightarrow 0$.
The two paradigmatic invisible-axion frameworks are the Dine-Fischler-Srednicki-Zhitnitsky~(DFSZ)~\cite{Zhitnitsky:1980tq,Dine:1981rt} and Kim-Shifman-Vainshtein-Zakharov~(KSVZ)~\cite{Kim:1979if,Shifman:1979if} scenarios. In DFSZ models, SM quarks are chirally charged under PQ requiring an extra Higgs doublet in its minimal realization, whereas in KSVZ models the anomaly is generated by additional heavy VLQs. In both scenarios PQ-symmetry breaking is done via a scalar singlet $\sigma$ allowing to decouple the PQ-breaking scale from the EW scale (for a review on axions see Ref.~\cite{DiLuzio:2020wdo}).
 
Axion frameworks also offer a gateway to address multiple beyond the SM phenomena simultaneously. In particular, axion relics can be generated non-thermally in the early Universe such that axions account for the observed DM abundance~\cite{Preskill:1982cy,Abbott:1982af,Dine:1982ah}. 
Moreover, the axion-neutrino connection has been extensively explored. Namely, in type-I seesaw~\cite{Minkowski:1977sc,Gell-Mann:1979vob,Yanagida:1979as,Schechter:1980gr,Glashow:1979nm,Mohapatra:1979ia}, where the VEV of a PQ-scalar $\sigma$, which sets the axion decay constant, also dynamically generates right-handed neutrino masses~\cite{Volkas:1988cm,Clarke:2015bea,Sopov:2022bog,RVolkas:2023jiv,Matlis:2023eli,Salvio:2015cja,Ballesteros:2016euj,Ballesteros:2016xej,Rocha:2025ade}. 
More recently, within the KSVZ framework, exotic colored vector-like fermions and scalar mediators have been employed to radiatively generate Majorana~\cite{Batra:2023erw} and Dirac~\cite{Batra:2025gzy} neutrino masses. Furthermore, the interplay between the PQ symmetry and the accidental $B$ and $L$ symmetries has been systematically studied in the DFSZ framework~\cite{Quevillon:2020hmx}. An axial version of the global $B+L$ symmetry has been identified with the PQ symmetry in a DFSZ axion scenario with tree-level proton decay~\cite{Reig:2018yfd}. The identification of $B+L$ with the PQ symmetry has also been studied in a KSVZ scenario, again with tree-level proton decay~\cite{Ohata:2021rkh}. Additionally, Ref.~\cite{Arias-Aragon:2022byr} studies UV completions of axion models with leptoquarks and diquarks, including various $B$ and $L$ breaking patterns, and analyzes tree-level nucleon decay with and without axion emission.

Inspired by these ideas, in this work we promote the accidental global $B+L$ symmetry of the SM to a PQ symmetry, connecting proton stability to the axion solution of the strong CP problem, while realizing spontaneous proton decay radiatively. The PQ anomaly sector, composed of VLQs, solves the strong CP problem within the KSVZ axion scenario. After spontaneous PQ breaking, a residual $\mathcal{Z}_2$ symmetry remains, forbidding proton decay at tree level. The VLQs responsible for the QCD anomaly, together with scalar mediators, odd under a residual $\mathcal{Z}_2$, induce proton decay at one-loop level. This paper is organized as follows. In Sec.~\ref{sec:framework}, we present the framework based on the symmetry-breaking pattern $\text{U}(1)_{\text{PQ}} \equiv \text{U}(1)_{B+L} \rightarrow \mathcal{Z}_2$, along with the particle content and their transformation properties under the SM gauge and PQ symmetries. We show how the proton decay operator $u_R u_R d_R e_R$ is generated spontaneously after PQ-breaking at one-loop and how the strong CP problem is addressed. In Sec.~\ref{sec:protondecaypheno}, we study proton decay phenomenology, focusing on two-body channels such as $p \rightarrow e^+ \pi^0$ and their correlation with the PQ-breaking scale. We also comment on suppressed processes involving an axion in the final state, such as $p \rightarrow e^+ \pi^0 a$. The different models can be experimentally distinguished through their predictions for the axion-to-photon coupling, analyzed in Sec.~\ref{sec:axionpheno}, where we also discuss axion DM production in the early Universe. Finally, our concluding remarks are presented in Sec.~\ref{sec:concl}.

%\newpage

%%%%%%%%%%%%%%%%%%%%%%%%%%%%%%%%%%%%%%%%%%%%%%%%%%%%%%%%%%%%%%%%%%%%%%%%%%%%%
\section{Framework}
\label{sec:framework}
%%%%%%%%%%%%%%%%%%%%%%%%%%%%%%%%%%%%%%%%%%%%%%%%%%%%%%%%%%%%%%%%%%%%%%%%%%%%%

%
\begin{table}[t!]
\renewcommand*{\arraystretch}{1.2}
	\centering
	\begin{tabular}{| c |c | c | c |}
		\hline 
&Fields&\SM&  U$(1)_{\text{PQ}} \equiv \text{U}(1)_{B+L}  \rightarrow \mathcal{Z}_2$\\
		\hline \hline
		\multirow{2}{*}{Leptons} 
&$\ell_L$&($\mathbf{1},\mathbf{2}, {-1/2}$)& $1 \rightarrow + $   \\
&$e_R$&($\mathbf{1},\mathbf{1}, {-1}$)& {$1 \rightarrow + $}    \\
\hline
\multirow{3}{*}{Quarks} 
&$q_L$&($\mathbf{3},\mathbf{2}, {1/6}$)& $1/3 \rightarrow + $   \\
&$d_R$&($\mathbf{3},\mathbf{1}, {-1/3}$)& {$1/3 \rightarrow +$}    \\
&$u_R$&($\mathbf{3},\mathbf{1}, {2/3}$)& {$1/3 \rightarrow +$} \\
\hline \hline
\multirow{2}{*}{Scalars} &$\Phi$&($\mathbf{1},\mathbf{2}, 1/2$)& {$0 \rightarrow +$}  \\
&$\sigma$&($\mathbf{1},\mathbf{1}, 0$)& {$1 \rightarrow +$}  \\	
\hline
	\end{tabular}
	\caption{Transformation properties under \SM $\;$ and U$(1)_\text{PQ}$, of the SM field content and $\sigma$ which breaks the PQ symmetry into a residual $\mathcal{Z}_2$ (see text for details).}
	\label{tab:B+LnuSM} 
\end{table}
We identify the accidental global $B+L$ symmetry already present in the SM, where quark and lepton fields have charge $1/3$ and $1$, respectively, with the PQ symmetry responsible for solving the strong CP problem, i.e. U(1)$_{\text{PQ}} \equiv \mathrm{U}(1)_{B+L}$. Besides the SM Higgs doublet $\Phi = \left(\phi^+, \phi^0 \right)^T$, whose VEV $\langle \phi^0 \rangle = v$, breaks the EW symmetry and has no charge under PQ, we introduce a complex scalar singlet $\sigma$ with charge~$1$ under U(1)$_{\text{PQ}}$. Once~$\sigma$ develops a non-zero VEV, the PQ symmetry is spontaneously broken at a scale $f_{\text{PQ}} = \left<\sigma \right>$, into a residual discrete $\mathcal{Z}_2$ symmetry. As shown in Table~\ref{tab:B+LnuSM}, the SM fields and $\sigma$ are even under $\mathcal{Z}_2$.

\begin{table}[t!]
\renewcommand*{\arraystretch}{1.2}
	\centering
	\begin{tabular}{| c |c | c | c | c | c | c | c|}
		\hline 
&\multirow{2}{*}{Fields}&\multirow{2}{*}{SU$(3)_{\text{c}}$} &\multirow{2}{*}{SU$(2)_{L}$} &\multicolumn{3}{c|}{U$(1)_{Y}$}&  \multirow{2}{*}{U$(1)_{\text{PQ}} \equiv \text{U}(1)_{B+L}  \rightarrow \mathcal{Z}_2$}\\
\cline{5-7}
&& & & Case A & Case B &  Case C&   \\
		\hline \hline
\multirow{2}{*}{Fermion} &$\Psi_{L};\Psi_{R}$&$\mathbf{R}_\Psi$ &$\mathbf{n}$ &$Y_\Psi$ &$Y_\Psi$ &$Y_\Psi$ & {$5/6; -1/6 \rightarrow - $}  \\ 
&$F_{L};F_{R}$&$\mathbf{R}_F$ & $\mathbf{n}$ & $-Y_\Psi-1/3$ & $-Y_\Psi-1/3$ & $-Y_\Psi-4/3$  & {$1/2; -1/2 \rightarrow -$}  \\ 
\hline \hline
\multirow{2}{*}{Scalar} &$S$&$\mathbf{R}_S$ &$\mathbf{n}$ &$-Y_\Psi+1/3$ & $-Y_\Psi-2/3$ & $-Y_\Psi-2/3$  & {$-1/6\rightarrow -$}   \\
&$\chi$&$\mathbf{R}_\chi$ &$\mathbf{n}$ &$Y_\Psi - 2/3$ & $Y_\Psi - 2/3$ &$Y_\Psi + 1/3$ & {$1/2 \rightarrow -$}   \\		
\hline
	\end{tabular}
	\caption{One-loop proton decay mediator fields $\Psi,F,S,\chi$, and their transformation properties under \SM $\;$ and U$(1)_\text{PQ}$. The field $\sigma$ presented in Table~\ref{tab:B+LnuSM}, breaks U(1)$_{\text{PQ}}$ into $\mathcal{Z}_2$. $\mathbf{R}_{\Psi,F,S,\chi}$ denote the SU(3)$_{\text{c}}$ representations of Table~\ref{tab:SU3Ncolor}. SU(2)$_L$ representation $\mathbf{n}$ is common to all fields. $Y_\Psi$ is the hypercharge of $\Psi$ with three possible cases A, B and C for hypercharge assignments (see text for details).}
	\label{tab:mediators} 
\end{table}
Furthermore, we introduce four fields consisting of two vector-like fermions, $\Psi$ and $F$, and two scalars, $S$ and $\chi$, with transformation properties under the SM gauge and PQ symmetries shown in Table~\ref{tab:mediators}. The fermions are chirally charged under $\mathrm{U}(1)_{\text{PQ}}$, so that
\begin{equation}
    \mathcal{L}_{\text{PQ}} = y_\Psi \overline{\Psi_L} \sigma \Psi_R + y_F \overline{F_L} \sigma F_R + \text{H.c.} \; ,
    \label{eq:LPQ}
\end{equation}
and after PQ-symmetry breaking their masses are given by
\begin{equation}
    M_{\Psi} = y_{\Psi} f_{\text{PQ}} \; , \; M_{F} = y_{F} f_{\text{PQ}} \; .
    \label{eq:FermionMasses}
\end{equation}
%
%$M_{\Psi,F} = y_{\Psi,F} f_{\text{PQ}}$. 
Note that the $\mathrm{SU}(2)_L$ representation $\mathbf{n} \equiv \mathbf{1}, \mathbf{2}, \mathbf{3}, \cdots$, is common to all fields. $Y_\Psi$ denotes the hypercharge of $\Psi$, and we consider three possible cases, A, B, and C, for the hypercharge assignments of the remaining fields. For the $\mathrm{SU}(3)_{\text{c}}$ representations $\mathbf{R}_{\Psi,F,S,\chi}$, Table~\ref{tab:SU3Ncolor} lists all possible color contractions up to $\mathbf{R} \equiv \mathbf{8}$. As shown in the last column of Table~\ref{tab:mediators}, the $\mathrm{U}(1)_{\text{PQ}}$ charge assignments of $\Psi$, $F$, $S$, and $\chi$ are such that these fields are odd under the residual $\mathcal{Z}_2$. For each case A, B, and C, the Yukawa Lagrangian contains the interactions
\begin{align}
    \mathcal{L}_{\text{A}} &\supset \mathbf{Y}_1 \overline{d_R^c} S \Psi_R + \mathbf{Y}_2 \overline{u_R} \chi^\dagger \Psi_L + \mathbf{Y}_3 \overline{u_R^c} S^\dagger F_R + \mathbf{Y}_4 \overline{e_R} \chi F_L + \text{H.c.} \; , \label{eq:LA} \\
    \mathcal{L}_{\text{B}} &\supset \mathbf{Y}_1 \overline{u_R^c} S \Psi_R + \mathbf{Y}_2 \overline{u_R} \chi^\dagger \Psi_L + \mathbf{Y}_3 \overline{d_R^c} S^\dagger F_R + \mathbf{Y}_4 \overline{e_R} \chi F_L + \text{H.c.} \; , \label{eq:LB} \\
    \mathcal{L}_{\text{C}} &\supset \mathbf{Y}_1 \overline{u_R^c} S \Psi_R + \mathbf{Y}_2 \overline{d_R} \chi^\dagger \Psi_L + \mathbf{Y}_3 \overline{u_R^c} S^\dagger F_R + \mathbf{Y}_4 \overline{e_R} \chi F_L + \text{H.c.} \; , \label{eq:LC}
\end{align}
where $\mathbf{Y}_{1,2,3,4}$ are $3 \times 1$ Yukawa vectors and for simplicity we omit the SU$(2)_L$ and SU$(3)_{\text{c}}$ indices. The most general scalar potential allowed by the symmetries of our framework can be written as
\begin{align}
V & = \sum_\zeta \left[\mu_{\zeta}^2 \; \zeta^{\dag}\zeta + \lambda_\zeta \left(\zeta^{\dag}\zeta\right)^2 \right] + \sum_{\zeta \neq \zeta^\prime}
\lambda_{\zeta \zeta^\prime}\left(\zeta^{\dag} \zeta\right)\left({\zeta^\prime}^{\dag} \zeta^\prime\right) \; ,
\label{eq:Vpotential}
\end{align}
where $\zeta, \zeta^\prime = \Phi, \sigma, S, \chi$.

\begin{figure}[!t]
    \centering
      \includegraphics[scale=0.85]{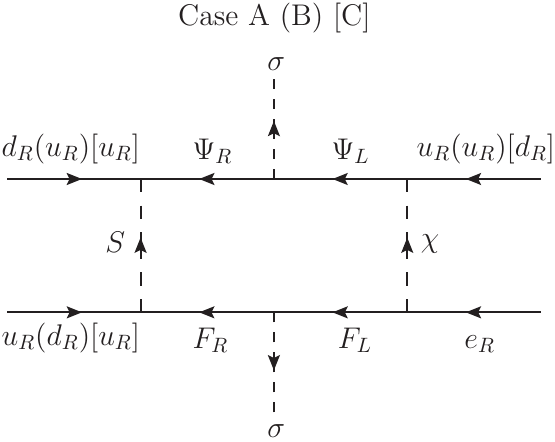}
      \caption{One-loop box diagram generating the proton decay operator $u_R u_R d_R e_R \sigma^{\ast 2}$ via the mediator fields $\Psi,F,S,\chi$ presented in Tables~\ref{tab:mediators} and~\ref{tab:SU3Ncolor} (see text for details).}
    \label{fig:Diagram_Operator}
\end{figure}

As shown in Ref.~\cite{Helo:2019yqp}, the fields $\Psi$, $F$, $S$, and $\chi$, mediate at one-loop level the dimension-six operator $u_R u_R d_R e_R$. However, since $\mathrm{U}(1)_{\text{PQ}}$ realizes $\mathrm{U}(1)_{B+L}$, the operator $u_R u_R d_R e_R$ is forbidden because it violates PQ or equivalently $B+L$, by two units. Instead, the operator $u_R u_R d_R e_R \sigma^{\ast 2}$ is generated at one-loop via the box diagram shown in Fig.~\ref{fig:Diagram_Operator}, with $u_R u_R d_R e_R$ being spontaneoulsy generated after PQ-symmetry breaking~\footnote{One-loop proton decay can also be generated through triangle diagrams, as discussed, e.g., in Ref.~\cite{Dorsner:2022twk}, via triple-leptoquark interactions.}. Additionally, once the breaking pattern $\mathrm{U}(1)_{\text{PQ}} \equiv \mathrm{U}(1)_{B+L} \rightarrow \mathcal{Z}_2$ occurs, all SM fields, including $u_R$, $d_R$, and $e_R$, as well as $\sigma$, are even under the residual $\mathcal{Z}_2$, whereas the mediator fields $\Psi$, $F$, $S$, and $\chi$ are odd (see last column of Table~\ref{tab:mediators}). Although the proton decay operator $u_R u_R d_R e_R$ is not forbidden by the residual $\mathcal{Z}_2$, as shown in Ref.~\cite{Kumar:2025aek} the even/odd character of the fields guarantees that proton decay cannot be generated at tree level~\footnote{See Ref.~\cite{Fonseca:2018ehk} for an example in which discrete $\mathcal{Z}_n$ symmetries are employed to forbid lower-dimensional proton decay operators.}. In Sec.~\ref{sec:protondecaypheno} we will study the phenomenology of spontaneous one-loop proton decay, which is connected to the PQ-symmetry breaking.

\begin{table}[t!]
	\centering
	\begin{tabular}{| c |c | c | c | c |}
		\hline 
$\mathbf{R}_\Psi$ & $\mathbf{R}_F$ & $\mathbf{R}_S$ &  $\mathbf{R}_\chi$ &  $N/n$ \\
		\hline \hline
		 $\mathbf{1}$ &$\mathbf{3}$&$\mathbf{\bar{3}}$& $\mathbf{\bar{3}}$ & 1 \\
        $\mathbf{3}$ &$\mathbf{1}$&$\mathbf{3}$& $\mathbf{1}$  & 1 \\ 
        \hline
        $\mathbf{\bar{3}}$ &$\mathbf{\bar{3}}$&$\mathbf{1}$& $\mathbf{3}$  & 2 \\
        $\mathbf{\bar{3}}$ &$\mathbf{\bar{3}}$&$\mathbf{8}$& $\mathbf{3}$  & 2 \\
        \hline
         $\mathbf{\bar{3}}$ &$\mathbf{6}$&$\mathbf{8}$& $\mathbf{\bar{6}}$ & 6  \\
        $\mathbf{6}$ &$\mathbf{\bar{3}}$&$\mathbf{8}$& $\mathbf{3}$  & 6 \\
        \hline
        $\mathbf{3}$ &$\mathbf{8}$&$\mathbf{3}$& $\mathbf{8}$  & 7 \\
        $\mathbf{3}$ &$\mathbf{8}$&$\mathbf{\bar{6}}$& $\mathbf{8}$  & 7 \\
        $\mathbf{8}$ &$\mathbf{3}$&$\mathbf{\bar{3}}$& $\mathbf{\bar{3}}$  & 7 \\
        $\mathbf{8}$ &$\mathbf{3}$&$\mathbf{3}$& $\mathbf{\bar{3}}$ & 7  \\
        \hline
        $\mathbf{\bar{6}}$ &$\mathbf{8}$&$\mathbf{3}$& $\mathbf{8}$  & 11 \\
         $\mathbf{8}$ &$\mathbf{\bar{6}}$&$\mathbf{\bar{3}}$& $\mathbf{6}$  & 11 \\
\hline
	\end{tabular}
	\caption{SU(3)$_{\text{c}}$ representations of the one-loop proton decay mediator fields $\Psi,F,S,\chi$ of Table~\ref{tab:mediators}. Last column: color anomaly factor $N/n$ values, where $n$ is the dimension of the SU(2)$_L$ representation [see Eq.~\eqref{eq:Nmodel}].}
	\label{tab:SU3Ncolor} 
\end{table}
The strong CP problem is solved within a KSVZ-type axion scenario since the vector-like fermions $\Psi$ and $F$ are chirally charged under the PQ symmetry as shown in Eq.~\eqref{eq:LPQ}. The PQ scalar field can be parametrized as,
\begin{equation}
    \sigma = (f_{\text{PQ}} + \rho) \exp\left(\frac{i a}{f_{\text{PQ}}}\right) \; ,
    \label{eq:sigmafield}
\end{equation}
where $a$ is the axion and $\rho$ is the radial mode. After spontaneous PQ breaking, the axion decay constant is
\begin{equation}
    f_a = \frac{f_{\text{PQ}}}{N} \; ,
    \label{eq:axionfacolor}
\end{equation}
where $N$ denotes the color anomaly factor. The up-to-date next-to-leading order (NLO) calculation of the axion mass yields~\cite{GrillidiCortona:2015jxo}
\begin{equation}
    m_a = 5.70(7) \left(\frac{10^{12} \text{GeV}}{f_a}\right) \mu \text{eV} \; .
    \label{eq:axionmass}
\end{equation}
This relation between $m_a$ and $f_a$ is a model-independent prediction of the QCD axion. Viability of the axion solution to the strong CP problem requires $N \neq 0$, ensuring an axion-gluon coupling. For the models in Tables~\ref{tab:mediators} and~\ref{tab:SU3Ncolor}, we obtain
\begin{align}
    N &= 2\, n \left[T(\mathbf{R}_\Psi) + T(\mathbf{R}_F)\right] \; ,
    \label{eq:Nmodel}
\end{align}
where $n$ is the dimension of the $\mathrm{SU}(2)_L$ representation $\mathbf{n}$ and $T(\mathbf{R}_{\Psi,F})$ is the Dynkin index of the $\mathrm{SU}(3)_{\text{c}}$ representation of $\Psi,F$. Hence, at least one of the vector-like fermions $\Psi$ or $F$ must transform nontrivially under $\mathrm{SU}(3)_{\text{c}}$ in order to solve the strong CP problem. In Table~\ref{tab:SU3Ncolor}, for the possible $\mathrm{SU}(3)_{\text{c}}$ representations $\mathbf{R}_{\Psi,F,S,\chi}$ up to $\mathbf{R}\equiv \mathbf{8}$, we report the corresponding values of $N/n$. Notice that in the cases $\mathbf{R}_{\Psi}\equiv \mathbf{1}$, $\mathbf{R}_{F}\equiv \mathbf{3}$ and $\mathbf{R}_{\Psi}\equiv \mathbf{3}$, $\mathbf{R}_{F}\equiv \mathbf{1}$, one of the vector-like fermions is a color singlet. In these situations, the color anomaly factor $N$ is entirely determined by the other fermion. These cases also yield $N=1$ if $\Psi$ and $F$ are $\mathrm{SU}(2)_L$ singlets, with implications for axion DM that we briefly discuss in Sec.~\ref{sec:axiondarkmatter}. All remaining cases feature a PQ anomaly sector with two VLQs, $\Psi$ and $F$, which also mediate proton decay at one loop, with the corresponding phenomenology discussed in the next section.

%%%%%%%%%%%%%%%%%%%%%%%%%%%%%%%%%%%%%%%%%%%%%%%%%%%%%%%%%%%%%%%%%%%%%%%%%%%%%
\section{Proton decay phenomenology}
\label{sec:protondecaypheno}
%%%%%%%%%%%%%%%%%%%%%%%%%%%%%%%%%%%%%%%%%%%%%%%%%%%%%%%%%%%%%%%%%%%%%%%%%%%%%

%
\begin{figure}[!t]
    \centering
      \includegraphics[scale=0.62]{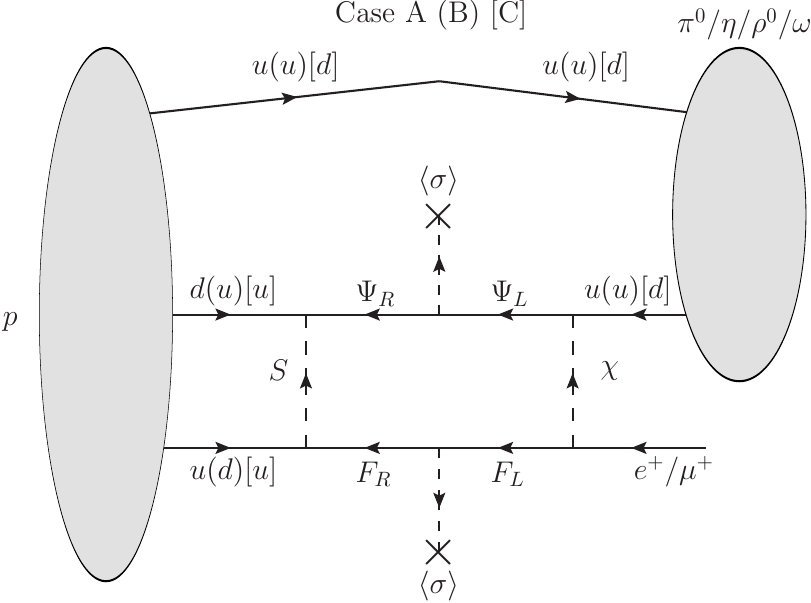} \hspace{+0.5cm} \includegraphics[scale=0.62]{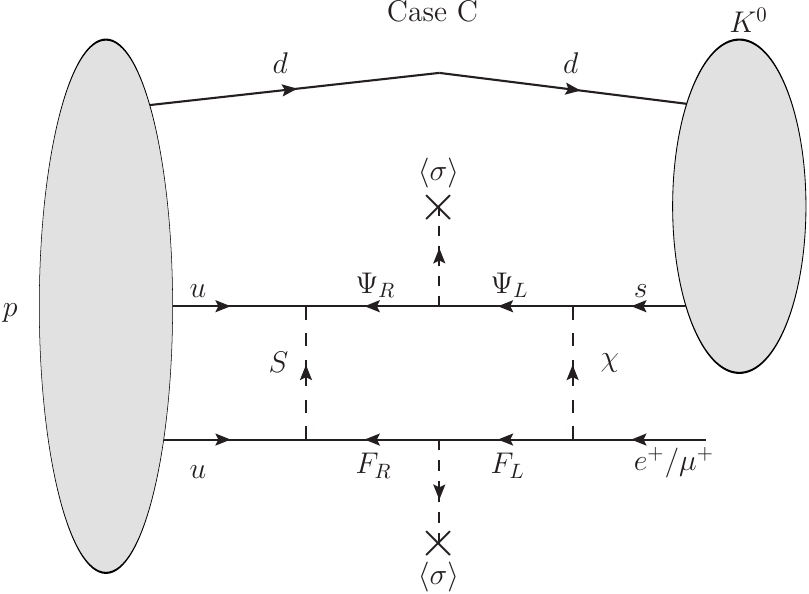} \\
      \vspace{+0.5cm}
      \includegraphics[scale=0.62]{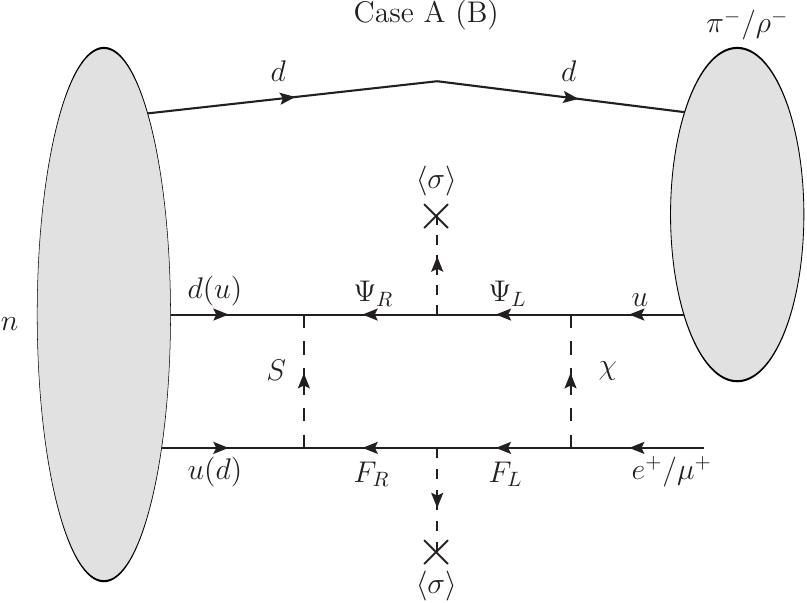}
      \caption{Examples of diagrams of one-loop proton (top) and neutron (bottom) two-body decays mediated by $\Psi,F,S$ and $\chi$ (see Fig.~\ref{fig:Diagram_Operator} and text for details).}
    \label{fig:Diagram_Proton_Decay}
\end{figure}
There are three possible hypercharge assignments for the fields $\Psi$, $F$, $S$, and $\chi$, denoted in Table~\ref{tab:mediators} as cases A, B, and C, which lead to distinct one-loop realizations of the operator $u_R u_R d_R e_R \sigma^{\ast 2}$, as illustrated in Fig.~\ref{fig:Diagram_Operator}. Once spontaneous breaking of the PQ symmetry is triggered by the VEV $\langle \sigma \rangle$, the dimension-six proton-decay operator $u_R u_R d_R e_R$ is generated. All three cases, A, B, and C, realize this operator at the one-loop level and consequently give rise to the same two-body proton decay modes, $p \rightarrow \ell^+ M^0$, where $\ell^+ = e^+, \mu^+$ and $M^0 = \pi^0, \eta, \rho^0, \omega, K^0$, as well as the neutron decay modes $n \rightarrow \ell^+ \pi^-$ and $n \rightarrow \ell^+ \rho^-$~\cite{Helo:2019yqp,Arbelaez:2026edo}. In Fig.~\ref{fig:Diagram_Proton_Decay}, we illustrate, as an example, the different one-loop two-body nucleon decay diagrams associated with cases A, B and C, when considering two of the three constituent quarks of the nucleon interact with the scalar $S$ in the loop.

\begin{table}[!t]
\renewcommand{\arraystretch}{1.5}
\centering
\begin{tabular}{|l|l|c|}   
\hline
Decay mode & Partial mean lifetime ($\times 10^{33}$ years) \\ 
\hline \hline
$p \rightarrow e^+ \pi^0 \; / \; \mu^+ \pi^0$ & $> 24 \; / \; 14$~\cite{Super-Kamiokande:2020wjk}  \\
$p \rightarrow e^+ \eta \; / \; \mu^+ \eta$ & $>14 \; / \;7.3$~\cite{Super-Kamiokande:2024qbv} \\
$p \rightarrow e^+ \rho^0 \; / \;  \mu^+ \rho^0$ & $>0.72 \; / \; 0.57$~\cite{Super-Kamiokande:2017gev} \\
$p \rightarrow e^+ \omega \; / \;  \mu^+ \omega$ & $>1.6 \; / \; 2.8$~\cite{Super-Kamiokande:2017gev} \\
$p \rightarrow e^+ K^0 \; / \;  \mu^+ K^0$ & $>1$~\cite{Super-Kamiokande:2017gev} $/ \; 3.6$~\cite{Super-Kamiokande:2022egr} \\
\hline \hline
$n \rightarrow e^+ \pi^- \; / \;  \mu^+ \pi^-$ & $>5.3 \; / \; 3.5$~\cite{Super-Kamiokande:2020wjk} \\
$n \rightarrow e^+ \rho^-\; / \; \mu^+ \rho^-$ & $>0.217 \; / \; 0.228$~\cite{Super-Kamiokande:2017gev} \\
\hline
\end{tabular}
\caption{Partial mean lifetime limit for proton and neutron two-body decay modes allowed in our framework (see Fig.~\ref{fig:Diagram_Proton_Decay}). All limits were obtained by the Super-K experiment at $90 \%$ CL.}
\label{tab:boundsProtonDecay}
\end{table}

\clearpage

In Table~\ref{tab:boundsProtonDecay}, we present the partial mean lifetime lower limits obtained by Super-K for the proton and neutron two-body decay modes allowed in our framework. Among them, the golden channel $p \rightarrow e^+ \pi^0$ provides the most stringent constraint, with the current lower bound on the partial lifetime given by $\tau\left(p \rightarrow e^+ \pi^0 \right) > 2.4 \times 10^{34}$ years~\cite{Super-Kamiokande:2020wjk}. Future searches at Hyper-K are expected to significantly improve the sensitivity to proton decay, potentially extending the reach for the $p \rightarrow e^+ \pi^0$ channel by about one order of magnitude in lifetime~\cite{Hyper-Kamiokande:2018ofw}. In what follows we focus on this paradigmatic proton decay channel and then comment on the other processes. 

As a benchmark scenario to study $p \rightarrow e^+ \pi^0$, and without loss of generality, we consider a concrete realization of our framework. In particular, we focus on case~A with $Y_\Psi = 0$ and, for simplicity, take the SU(2)$_L$ representation to be $\mathbf{n} \equiv \mathbf{1}$ (see Table~\ref{tab:mediators}). For the SU(3)$_{\text{c}}$ representations, we likewise consider the simplest scenario involving only two colored mediators, namely $\mathbf{R}_\Psi \equiv \mathbf{3}$ and $\mathbf{R}_F \equiv \mathbf{1}$ for the fermionic sector, together with $\mathbf{R}_S \equiv \mathbf{3}$ and $\mathbf{R}_\chi \equiv \mathbf{1}$ for the scalar sector (see Table~\ref{tab:SU3Ncolor}). The vector-like fermion masses are dynamically generated through PQ-symmetry breaking, as given in Eq.~\eqref{eq:FermionMasses}. Furthermore, from the scalar potential in Eq.~\eqref{eq:Vpotential}, the scalar masses of $S$ and $\chi$ are obtained as
\begin{equation}
    M^2_{S} = \mu_{S}^2 + \lambda_{\Phi S} v^2 + \lambda_{\sigma S} f_{\text{PQ}}^2 \; , \; M^2_{\chi} = \mu_{\chi}^2 + \lambda_{\Phi \chi} v^2 + \lambda_{\sigma \chi} f_{\text{PQ}}^2 \; .
    \label{eq:ScalarMasses}
\end{equation}
Since both $S$ and $\chi$ are electrically charged, and $S$ is additionally colored, neither field can acquire a VEV in order to preserve electromagnetism and color. This requirement implies that the quadratic parameters must satisfy $\mu_{S,\chi}^2 > 0$. 

The proton decay width induced at one-loop level for the two-body mode $p \rightarrow e^+ \pi^0$ can be written as
\begin{equation}
\Gamma(p\rightarrow e^+\pi^0)=
\frac{m_p}{32\pi}
\left(1-\frac{m_{\pi^0}^2}{m_p^2}\right)^2
\left|W_0\,\mathcal{C}\right|^2 \; ,
\label{eq:ProtonDecayWidth}
\end{equation}
with the proton and neutral pion masses being $m_p = 0.938$ GeV and $m_{\pi^0} = 0.135$ GeV, respectively~\cite{ParticleDataGroup:2024cfk}. Since we radiatively generate the operator $u_R u_R d_R e_R \sigma^{\ast 2}$, the value of the hadronic matrix element above is $W_0 = -0.131~\text{GeV}^2$~\cite{Aoki:2017puj}. The loop factor $\mathcal{C}$ is~\cite{Helo:2019yqp},
\begin{equation}
\mathcal{C} = -|Y_1 Y_2 Y_3 Y_4| \; M_\Psi M_F \; \mathcal{I}(M_\Psi,M_F,M_S,M_\chi) \; ,
\label{eq:Cdef}
\end{equation}
where the Yukawa couplings $Y_{1,2,3,4}$ correspond to the first-generation fermion entries of the Yukawa vectors $\mathbf{Y}_{1,2,3,4}$ of Eq.~\eqref{eq:LA}. The box-diagram loop integral $\mathcal{I}$ (see Figs.~\ref{fig:Diagram_Operator} and~\ref{fig:Diagram_Proton_Decay}), is a function of the mediator masses $M_\Psi,M_F$ and $M_S,M_\chi$, given by Eqs.~\eqref{eq:FermionMasses} and~\eqref{eq:ScalarMasses}, respectively. The proton decay partial mean lifetime is obtained through,
\begin{equation}
\tau(p\rightarrow e^+\pi^0) = \frac{1}{\Gamma(p\rightarrow e^+\pi^0)} \; .
\label{eq:ProtonLifetime}
\end{equation}

The framework considered here is an invisible KSVZ-axion scenario, where the PQ symmetry is identified as $B+L$. Once the PQ-symmetry, and therefore $B+L$, is spontaneously broken one-loop proton decay is generated. The PQ-breaking scale $f_{\text{PQ}} = \langle \sigma \rangle$ is typically of order $10^{12}$ GeV [see Eq.~\eqref{eq:axionmass}], and thus lies many orders of magnitude above the EW scale, $f_{\text{PQ}} \gg v \simeq 246$ GeV. As will be discussed in Sec.~\ref{sec:axiondarkmatter} and shown in Fig.~\ref{fig:gagg}, such a PQ scale naturally allows axions to account for the observed DM relic abundance. Consequently, due to the connection between PQ-symmetry breaking and $B+L$ violation, the proton decay mediator fields $\Psi$, $F$, $S$, and $\chi$ are expected to acquire masses of order $f_{\text{PQ}}$ [see Eqs.~\eqref{eq:FermionMasses} and~\eqref{eq:ScalarMasses}]. In the limit where all masses in the loop are heavy compared to external momenta and identical $M_\Psi=M_F=M_S=M_\chi\equiv M$, the loop integral is given by,
\begin{align}
\mathcal{I}(M) &\simeq \frac{1}{16\pi^2} \frac{1}{6 M^4} \; .
\end{align}
In this limit we obtain the parametric expression for the proton decay lifetime
\begin{equation}
\tau(p\rightarrow e^+\pi^0) \simeq 10^{34} \; \text{years} \; \left(\frac{10^{-4}}{Y_1 Y_2 Y_3 Y_4}\right)^2 \; \left(\frac{M}{10^{12} \; \text{GeV}}\right)^4 \; .
\label{eq:ProtonDecayWidth}
\end{equation}
The above expression is a reasonable approximation since $M_\Psi \sim M_F \sim M_S \sim M_\chi \sim \mathcal{O}(f_{\text{PQ}}) \gg v$, for $y_{\Psi,F} \sim \mathcal{O}(1)$ and $\lambda_{\sigma S,\chi} \sim \mathcal{O}(1)$. We note that, in order to satisfy the Super-K constraint shown in Table~\ref{tab:boundsProtonDecay}, for $M \simeq f_{\text{PQ}} \simeq 10^{12}$ GeV, the Yukawa combination must satisfy $Y_1 Y_2 Y_3 Y_4 \lesssim 10^{-4}$. Therefore, taking all aforementioned Yukawa and scalar quartic couplings to lie in the range $\mathcal{O}(0.1\text{--}1)$ easily satisfies the stringent proton decay bounds at the typical PQ scale of $10^{12}$ GeV. This follows from the intimate connection between PQ-symmetry breaking and $B+L$ violation, together with the fact that proton decay is generated at one loop, leading to an additional suppression of the decay rate and consequently an enhancement of the corresponding lifetime. Note that as in Refs.~\cite{Helo:2019yqp,Kumar:2025aek}, suppressing the Yukawa combination $Y_1 Y_2 Y_3 Y_4$ down to values below $10^{-20}$, which corresponds to individual couplings smaller than $10^{-5}$, would allow for the mediator masses to lie at the TeV scale, which can lead to interesting phenomenological implications such as viable WIMP DM candidates and testable collider signatures. In our framework, however, DM is provided by the axion (see Sec.~\ref{sec:axiondarkmatter}) and the VLQs and scalar leptoquarks are expected to lie far beyond the direct reach of collider experiments. Lowering their masses to the TeV scale, so as to make them accessible at colliders, would require unnaturally small Yukawa couplings, $y_{\Psi,F} \sim \mathcal{O}(10^{-9})$, and/or scalar quartic couplings, $\lambda_{\sigma S,\chi} \sim \mathcal{O}(10^{-18})$. Nevertheless, as we will show in Sec.~\ref{sec:axionphoton}, the different mediator field representations listed in Tables~\ref{tab:mediators} and~\ref{tab:SU3Ncolor} can still be distinguished indirectly through their distinct predictions for the axion-to-photon coupling.

Regarding the other decay channels illustrated in Fig.~\ref{fig:Diagram_Proton_Decay}, which are currently less constrained than $p \rightarrow e^+ \pi^0$ as indicated in Table~\ref{tab:boundsProtonDecay}, a detailed analysis would lead to conclusions qualitatively similar to those obtained for the $p \rightarrow e^+ \pi^0$ channel. Nevertheless, it is important to emphasize that some of these decay modes provide complementary probes of the same underlying interactions, while others are sensitive to additional Yukawa couplings. In particular, the proton decay channels $p \rightarrow e^+ \eta$, $p \rightarrow e^+ \rho^0$, and $p \rightarrow e^+ \omega$, as well as, the neutron decay modes $n \rightarrow e^+ \pi^-$ and $n \rightarrow e^+ \rho^-$, probe the same combination of Yukawa couplings $Y_{1,2,3,4}$ as $p \rightarrow e^+ \pi^0$, corresponding to the first-generation fermion entries of the Yukawa vectors $\mathbf{Y}_{1,2,3,4}$ appearing in Eqs.~\eqref{eq:LA},~\eqref{eq:LB}, and~\eqref{eq:LC}. On the other hand, channels involving $\mu^+$ are sensitive to the second-generation entry of the Yukawa vector $\mathbf{Y}_{4}$. Furthermore, the decay channels $p \rightarrow e^+ K^0$ and $p \rightarrow \mu^+ K^0$, probe the second-generation entry of the Yukawa vector $\mathbf{Y}_{1}$, $\mathbf{Y}_{3}$ and $\mathbf{Y}_{2}$, for cases A, B and C, respectively.

Lastly, since PQ-symmetry breaking lies at the origin of $B+L$ violation, triggering spontaneous one-loop proton decay, our framework naturally predicts exotic proton decay channels with an axion in final state, such as $p \rightarrow e^+ \pi^0 a$ and $p \rightarrow e^+ \pi^0 a a$. Namely, using Eq.~\eqref{eq:sigmafield} and expanding the $\sigma$ field, we obtain
\begin{equation}
\frac{\mathcal{C}}{f_\text{PQ}^2} u_R u_R d_R e_R \sigma^{\ast 2}
\; \rightarrow \;
\mathcal{C}
u_R u_R d_R e_R
\left(
1
-\frac{2 i a}{f_{\text{PQ}}}
-\frac{2 a^2}{f_{\text{PQ}}^2}
+\cdots
\right) \; ,
\end{equation}
where the same loop coefficient $\mathcal{C}$ defined in Eq.~\eqref{eq:Cdef}, which controls the standard two-body decay $p \rightarrow e^+ \pi^0$, also determines the effective couplings associated with one and two axion emission. Although the axion appears here through non-derivative interactions, this description is equivalent, after a standard PQ rotation of the fermion fields, to the usual derivative coupling of the axion to fermions as a pseudo-Goldstone boson~\footnote{For comprehensive studies of baryon-number-violating decays involving axions and axion-like particles from an effective field theory perspective, see Refs.~\cite{Li:2024liy,Fan:2025xhi}.}. Consequently, axion-emission amplitudes are effectively controlled by the characteristic momentum transfer of the process, namely $p \sim m_p$. For the typical invisible-axion scale $f_{\text{PQ}} \sim \mathcal{O}(10^{12})$ GeV, these exotic channels are therefore expected to be extremely suppressed. In particular, the decay widths satisfy the hierarchy
\begin{equation}
\frac{\Gamma(p\rightarrow e^+\pi^0 \; a \; a)}{\Gamma(p\rightarrow e^+\pi^0 \; a)}
\sim
\frac{\Gamma(p\rightarrow e^+\pi^0 \; a)}{\Gamma(p\rightarrow e^+\pi^0)}
\sim
\frac{1}{16\pi^2} \left(\frac{m_p}{f_{\text{PQ}}}\right)^2
\sim
\mathcal{O}(10^{-27})
\ll 1 \; ,
\end{equation}
% \frac{m_p^2}{16\pi^2 f_{\text{PQ}}^2}
where each emitted axion introduces an additional suppression factor $\sim m_p/f_{\text{PQ}}$ in the amplitude, while the multi-body final states are further suppressed by the usual phase-space factor $\sim 1/16\pi^2$. Consequently, from the above it is clear that the standard two-body proton decay modes dominate over the channels with axions in the final state.

%%%%%%%%%%%%%%%%%%%%%%%%%%%%%%%%%%%%%%%%%%%%%%%%%%%%%%%%%%%%%%%%%%%%%%%%%%%%%
\section{Axion phenomenology}
\label{sec:axionpheno}
%%%%%%%%%%%%%%%%%%%%%%%%%%%%%%%%%%%%%%%%%%%%%%%%%%%%%%%%%%%%%%%%%%%%%%%%%%%%%

%%%%%%%%%%%%%%%%%%%%%%%%%%%%%%%%%%%%%%%%%%%%%%%%%%%%%%%%%%%%%%%%%%%%%%%%%%%%%
\subsection{Axion dark matter and cosmology}
\label{sec:axiondarkmatter}
%%%%%%%%%%%%%%%%%%%%%%%%%%%%%%%%%%%%%%%%%%%%%%%%%%%%%%%%%%%%%%%%%%%%%%%%%%%%%

Axions are naturally light, weakly coupled to ordinary matter, nearly stable, and can be non-thermally produced in the early Universe, providing an excellent DM candidate. Axion DM production depends on whether the PQ symmetry breaks before or after inflation. We briefly discuss both possibilities below.

In the pre-inflationary scenario, the PQ symmetry is broken before inflation. In this case, axion DM is generated entirely through the misalignment mechanism~\cite{Preskill:1982cy,Abbott:1982af,Dine:1982ah}. The relic axion abundance can be approximately written as~\cite{DiLuzio:2020wdo} 
\begin{equation}
\Omega_a h^2 \simeq  \Omega_\text{DM} h^2 \left(\frac{\theta_0}{2.15}\right)^2 \left(\frac{f_a}{2 \times 10^{11} \ \text{GeV}} \right)^{\frac{7}{6}} \; ,
\label{eq:relica}
\end{equation}
where $|\theta_0| \in [0, \pi)$ denotes the initial misalignment angle and the observed cold DM relic abundance, obtained by Planck satellite data is $\Omega_{\text{DM}} h^2 = 0.1200 \pm 0.0012$~\cite{Planck:2018vyg}. Hence, for $0.1<\theta_0<\pi$ the axion decay constant needs to be $10^{11} \ \text{GeV} <f_a<4\times 10^{13} \ \text{GeV}$, in order for axions to account for all the observed DM relic abundance. As discussed in Sec.~\ref{sec:axionphoton} and shown in Fig.~\ref{fig:gagg}, this parameter space region is currently being probed by haloscope experiments. In this scenario, the axion leaves an imprint in primordial fluctuations, reflected in the cosmic microwave background~(CMB) anisotropies and large-scale structure. The resulting isocurvature fluctuations are constrained by CMB data~\cite{Beltran:2006sq}, leading to an upper bound on the inflationary scale $H_I$~\cite{DiLuzio:2016sbl}:
\begin{equation}
   H_I \lsim  \frac{0.9\times 10^7}{\Omega_a h^2/\Omega_\text{DM} h^2} \left(\frac{\theta_0}{\pi} \frac{f_a}{ 10^{11} \ \text{GeV}} \right) \; \text{GeV} \; .
   \label{eq:Iso}
\end{equation}
Taking $\theta_0 \sim \mathcal{O}(1)$ and $\Omega_a h^2=0.12$, leads to a low scale for inflation $H_I \lsim 10^7$ GeV, with Planck currently probing $H_I \lsim 10^{13}$ GeV~\cite{Planck:2018vyg}. Nonetheless, the isocurvature bound depends on the UV completion and on the specific inflationary mechanism at play, and can therefore be relaxed in concrete scenarios~\cite{Graham:2025iwx}. In this work, however, we do not specify an explicit inflationary setup in which this CMB constraint can be consistently assessed. A dedicated study of inflation in such KFSZ models is left for future work, along the lines of Refs.~\cite{Ballesteros:2016euj,Ballesteros:2016xej}.

In the post-inflationary scenario, the PQ symmetry is broken after inflation, resulting in an observable Universe composed of patches with different initial values of the axion field. Consequently, the initial misalignment angle $\theta_0$ is no longer a free parameter, but is instead statistically determined, with the current estimate $\langle \theta_0^2 \rangle \simeq 2.15^2$~\cite{DiLuzio:2020wdo}. This makes the scenario, in principle, more predictive than the pre-inflationary case. Requiring $\Omega_a h^2 = \Omega_{\text{DM}} h^2$ then fixes the axion scale $f_a$ (or equivalently $m_a$) [see Eq.~\eqref{eq:relica}]. If the axion abundance arises solely from the misalignment mechanism, avoiding DM overproduction implies $f_a \lesssim 2 \times 10^{11}$~GeV. However, the situation is more involved since topological defects, namely axion strings and domain walls~(DWs), also contribute to the total relic abundance $\Omega_a h^2$~\cite{Bennett:1987vf,Levkov:2018kau,Gorghetto:2018myk,Buschmann:2019icd}. The non-linear evolution of the resulting DW-string network must be studied numerically, and its precise contribution remains an open question. A detailed analysis lies beyond the scope of this work. Nonetheless, the cosmological DW problem is absent for $N_{\text{DW}}= N = 1$. $N_{\text{DW}}$ denotes the vacuum degeneracy of the axion potential associated with the residual $\mathcal{Z}_N$ symmetry left unbroken by QCD instanton effects that anomalously break U(1)$_{\text{PQ}}$. This occurs for the models in Tables~\ref{tab:mediators} and~\ref{tab:SU3Ncolor}, where $\Psi$ and $F$ are SU(2)$_L$ singlets, i.e. $\mathbf{n}=1$, with their SU(3)$_{\text c}$ representations consisting of one being a color triplet $\mathbf{3}$ while the other is a color singlet $\mathbf{1}$. All other cases lead to DW formation in the early Universe~\cite{Lazarides:2018aev}. Even for $N_{\text{DW}}=1$, axionic string networks can still form. Numerical simulations indicate that reproducing the observed DM abundance, $\Omega_a h^2 = \Omega_{\text{DM}} h^2$, requires $f_a \in [5 \times 10^9 , 3 \times 10^{11}]$~GeV~\cite{Kawasaki:2014sqa,Klaer:2017ond,Gorghetto:2020qws,Buschmann:2021sdq,Benabou:2024msj}. Even in scenarios with $N_{\text{DW}}>1$, one may introduce small bias terms in the scalar potential that explicitly break the residual $\mathcal{Z}_N$ symmetry of the axion potential. This lifts the vacuum degeneracy and renders the DW-string network unstable~\cite{Sikivie:1982qv}. The subsequent decay of strings and DWs may then generate potentially observable stochastic gravitational wave signals~\cite{Roshan:2024qnv,Servant:2023mwt,Morais:2023ciz}.

\newpage

The one-loop proton decay mediators $\Psi$, $F$, $S$, and $\chi$ are odd under $\mathcal{Z}_2$. As a consequence, the lightest $\mathcal{Z}_2$-odd state is stable and can be thermally produced after inflation. This may lead to cosmological problems, particularly if the lightest state among $\Psi$, $F$, $S$, and $\chi$ is colored and/or electrically charged~\cite{Nardi:1990ku,Perl:2001xi,Perl:2009zz}. Indeed, searches in terrestrial, lunar, and meteoritic materials place stringent bounds on such stable baryonic and/or charged relics~\cite{Perl:2001xi,Perl:2009zz,Burdin:2014xma,Hertzberg:2016jie,Mack:2007xj}, excluding them unless some mechanism suppresses their abundance or allows them to decay into ordinary matter~\cite{Nardi:1990ku,DiLuzio:2016sbl,DiLuzio:2017pfr}. Among the mediator representations listed in Tables~\ref{tab:mediators} and~\ref{tab:SU3Ncolor}, the problematic stable relics can be avoided in scenarios where either $\Psi$ or $F$ is a fermionic SM singlet, or where $S$ or $\chi$ is a scalar SM singlet. In fact, Ref.~\cite{Kumar:2025aek} exploits a global $B+L$ symmetry broken to a residual $\mathcal{Z}_4$ symmetry to stabilize a viable WIMP DM candidate corresponding to either a fermionic or scalar singlet mediator participating in one-loop proton decay. In our framework, taking $\Psi$, $F$, $S$, and $\chi$ to be SU(2)$_L$ singlets allows for scenarios containing either one fermionic or scalar SM singlet that also feature $N_{\text{DW}}=1$. These cosmologically conservative models are more predictive since post-inflationary axion DM, free of DWs, becomes viable. Namely, these models correspond to:
\begin{itemize}
    \item \textbf{Models 1 (M$_1$):} The fermionic singlet scenario, $\Psi \sim \left(\mathbf{1},\mathbf{1},0\right)$ is obtained for $Y_\Psi = 0$ together with $\mathbf{R}_\Psi \equiv \mathbf{1}$. As shown in Table~\ref{tab:SU3Ncolor}, this implies $\mathbf{R}_F \equiv \mathbf{3}$ and $\mathbf{R}_S = \mathbf{R}_\chi \equiv \mathbf{\bar{3}}$.
    \item \textbf{Models 2 (M$_2$):} Alternatively, $F \sim \left(\mathbf{1},\mathbf{1},0\right)$ is realized by taking $Y_\Psi = -1/3$ for cases A and B, or $Y_\Psi = -4/3$ for case C, together with $\mathbf{R}_F \equiv \mathbf{1}$. This implies $\mathbf{R}_\Psi = \mathbf{R}_S \equiv \mathbf{3}$ and $\mathbf{R}_\chi \equiv \mathbf{1}$.
    \item \textbf{Models 3 (M$_3$):} The scalar singlet scenario, $\chi \sim \left(\mathbf{1},\mathbf{1},0\right)$ is obtained by taking $Y_\Psi = 2/3$ for cases A and B, or $Y_\Psi = -1/3$ for case C, together with $\mathbf{R}_\chi \equiv \mathbf{1}$. In this case $\mathbf{R}_\Psi = \mathbf{R}_S \equiv \mathbf{3}$ and $\mathbf{R}_F \equiv \mathbf{1}$.
\end{itemize}
By contrast, the $S\sim (\mathbf{1}, \mathbf{1}, 0)$ scenario leads to $N_{\text{DW}}=2$ (see Table~\ref{tab:SU3Ncolor}), and therefore inevitably suffers from the cosmological DW problem. Hence, apart from the highlighted models \textbf{M}$_{1,2,3}$ above, for the majority of models within our framework, it is therefore judicious to assume a pre-inflationary axion DM scenario. In fact, since the breaking of U(1)$_{\text{PQ}}$ occurs before inflation, axionic strings, DWs, and any primordial abundance of stable baryonic and/or charged relics will be washed out during inflation.

%%%%%%%%%%%%%%%%%%%%%%%%%%%%%%%%%%%%%%%%%%%%%%%%%%%%%%%%%%%%%%%%%%%%%%%%%%%%%
\subsection{Axion-to-photon coupling}
\label{sec:axionphoton}
%%%%%%%%%%%%%%%%%%%%%%%%%%%%%%%%%%%%%%%%%%%%%%%%%%%%%%%%%%%%%%%%%%%%%%%%%%%%%

%
\begin{table}[t!]
\renewcommand*{\arraystretch}{1.2}
\centering
\begin{tabular}{|c|c|c|c|c|}
\hline
\multicolumn{5}{|c|}{Case A/B (Case C): $E/N$ values for $Y_\Psi = 0$} \\
\hline \hline
\multicolumn{2}{|c|}{SU($3)_\text{c}$} & \multicolumn{3}{c|}{SU($2)_L$} \\
\hline
$\mathbf{R}_\Psi$ & $\mathbf{R}_F$ & $\mathbf{1}$ & $\mathbf{2}$ & $\mathbf{3}$ \\
\hline
$\mathbf{1}$ & $\mathbf{3}$ 
& $2/3\,(32/3)$ 
& $8/3\,(38/3)$ 
& $6\,(16)$ \\

$\mathbf{3}$ & $\mathbf{1}$ 
& $2/9\,(32/9)$ 
& $20/9\,(50/9)$ 
& $50/9\,(80/9)$ \\

$\mathbf{\bar{3}}$ & $\mathbf{\bar{3}}$ 
& $1/3\,(16/3)$ 
& $11/6\,(41/6)$ 
& $13/3\,(28/3)$ \\

$\mathbf{\bar{3}}$ & $\mathbf{6}$ 
& $2/9\,(32/9)$ 
& $35/36\,(155/36)$ 
& $20/9\,(50/9)$ \\

$\mathbf{6}$ & $\mathbf{\bar{3}}$ 
& $1/9\,(16/9)$ 
& $31/36\,(91/36)$ 
& $19/9\,(34/9)$ \\

$\mathbf{3}$ & $\mathbf{8}$ 
& $16/63\,(256/63)$ 
& $131/126\,(611/126)$ 
& $148/63\,(388/63)$ \\

$\mathbf{8}$ & $\mathbf{3}$ 
& $2/21\,(32/21)$ 
& $37/42\,(97/42)$ 
& $46/21\,(76/21)$ \\

$\mathbf{\bar{6}}$ & $\mathbf{8}$ 
& $16/99\,(256/99)$ 
& $79/99\,(29/9)$ 
& $184/99\,(424/99)$ \\

$\mathbf{8}$ & $\mathbf{\bar{6}}$ 
& $4/33\,(64/33)$ 
& $25/33\,(85/33)$ 
& $20/11\,(40/11)$ \\
\hline
\end{tabular}
\caption{$E/N$ values for $Y_\Psi = 0$, for Case A/B (Case C) -- see Tables~\ref{tab:mediators} and~\ref{tab:SU3Ncolor} and Eqs.~\eqref{eq:ENmodelAB} and~\eqref{eq:ENmodelC}.}
\label{tab:EN}
\end{table}
The various models presented in Tables~\ref{tab:mediators} and~\ref{tab:SU3Ncolor} can be probed experimentally via their predictions for the axion-to-photon coupling $g_{a \gamma \gamma}$. NLO chiral Lagrangian techniques provide the following result~\cite{GrillidiCortona:2015jxo}~\footnote{See Ref.~\cite{Brandt:2026wir} for a determination of the axion-to-photon coupling via lattice QCD.} 
\begin{align}
g_{a \gamma \gamma} &= \frac{\alpha_e}{2 \pi f_a} \left[\frac{E}{N} - 1.92(4) \right] \; ,
\label{eq:gagg}
\end{align}
where $\alpha_e = e^2/(4\pi)$, and the ratio $E/N$ is the model-dependent contribution, with $E$ being the electromagnetic anomaly factor.

%\newpage

As shown in Table~\ref{tab:mediators}, the allowed hypercharge assignments for the one-loop proton-decay mediators $\Psi$, $F$, $S$, and $\chi$ fall into three classes, labelled A, B, and C. Cases A and B yield identical hypercharges for the vector-like fermions $\Psi$ and $F$, whereas case C leads to a distinct assignment. Moreover, since all mediators share the same $\mathrm{SU}(2)_L$ representation, there are only two independent expressions for $E$, namely
\begin{align}
E_{\text{A/B}} &= 2 \sum_{j=0}^{n-1} \left[ d(\mathbf{R}_\Psi) \left(j - \frac{n-1}{2} + Y_\Psi\right)^2
+ d(\mathbf{R}_F)\left(j - \frac{n-1}{2} - Y_\Psi - \frac{1}{3}\right)^2 \right] \, ,
\label{eq:ENmodelAB} \\
E_{\text{C}} &= 2 \sum_{j=0}^{n-1} \left[ d(\mathbf{R}_\Psi) \left(j - \frac{n-1}{2} + Y_\Psi\right)^2
+ d(\mathbf{R}_F)\left(j - \frac{n-1}{2} - Y_\Psi - \frac{4}{3}\right)^2 \right] \, ,
\label{eq:ENmodelC}
\end{align}
where $d(\mathbf{R}_{\Psi,F})$ is the dimension of the $\mathrm{SU}(3)_c$ representation of $\Psi,F$. The quantities in parentheses correspond to the electric charges of the individual components of the $\mathrm{SU}(2)_L$ multiplets $\Psi$ and $F$. Consequently, for fixed $Y_\Psi$ and $\mathbf{n}$, cases A and B give identical model-dependent contributions to $g_{a\gamma\gamma}$. In Table~\ref{tab:EN} we list the resulting $E/N$ values for $Y_\Psi=0$, entries are given for case A/B, with the corresponding case C values shown in parentheses. We include all allowed $\mathrm{SU}(3)_c$ representations of $\Psi$ and $F$ shown in Table~\ref{tab:SU3Ncolor}, and consider $\mathrm{SU}(2)_L$ multiplets up to $\mathbf{n}\equiv3$. 

\begin{figure}[!t]
    \centering
      \includegraphics[scale=0.122]{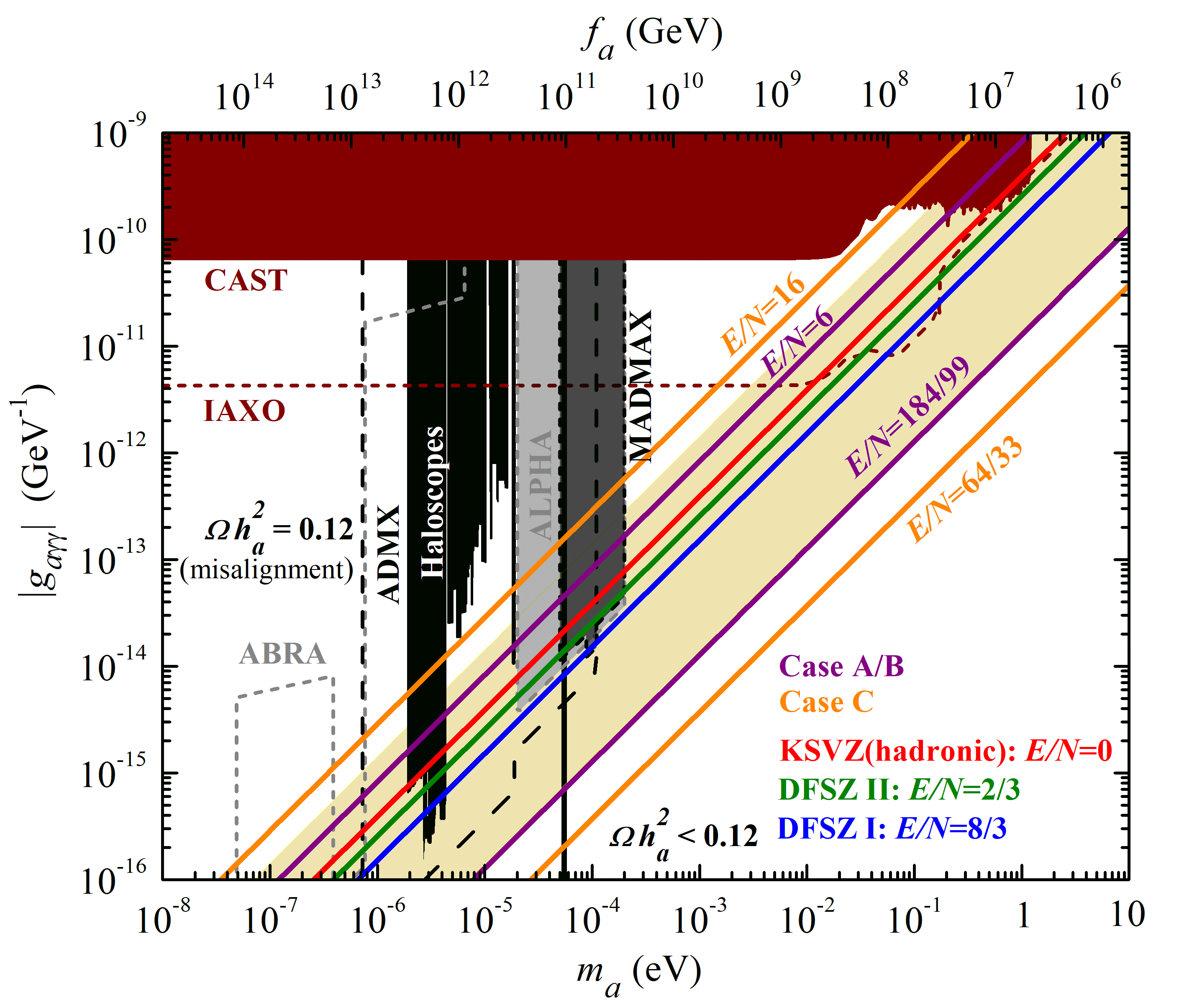} \hspace{-0.4cm} \includegraphics[scale=0.122]{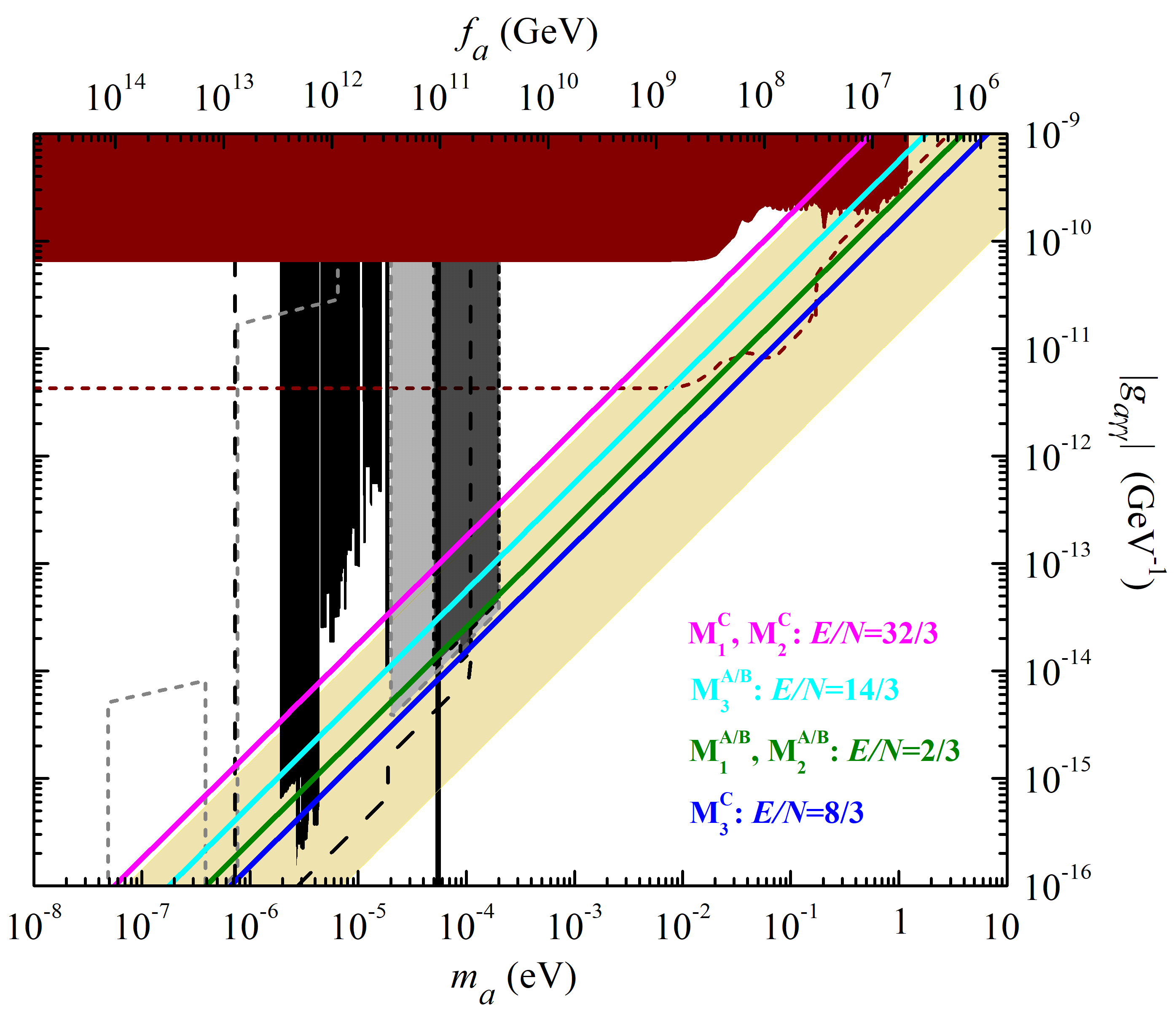}
      \caption{Axion-to-photon coupling $|g_{a \gamma \gamma}|$ versus axion mass $m_a$ (bottom axis) and decay constant $f_a$ (top axis). Left:~Predictions for the models of Table~\ref{tab:EN}, where the purple (orange) lines correspond to $E/N$ values leading to maximum and minimum $|g_{a \gamma \gamma}|$ for case A/B (C). The KSVZ (hadronic) and DFSZ I and II predictions are indicated by the red, blue, and green lines, respectively. Right: Predictions for the cosmologically conservative models \textbf{M}$_{1,2,3}$ identified in Sec.~\ref{sec:axiondarkmatter}, are indicated by the green, blue, cyan and magenta lines. The yellow shaded region refers to the usual QCD axion window~\cite{DiLuzio:2016sbl}. Current constraints from helioscopes like CAST~\cite{CAST:2017uph} exclude the burgundy shaded region, while haloscopes like ADMX~\cite{ADMX:2018gho,ADMX:2019uok,ADMX:2021nhd}, RBF~\cite{DePanfilis:1987dk}, CAPP~\cite{CAPP:2020utb} and HAYSTAC~\cite{HAYSTAC:2020kwv}, rule out the black region. Projected sensitivities of IAXO~\cite{Shilon_2013}, ADMX~\cite{Stern:2016bbw}, MADMAX~\cite{Beurthey:2020yuq}, ALPHA~\cite{ALPHA:2022rxj} and ABRACADABRA~\cite{Ouellet:2018beu} are indicated by the dashed burgundy, dashed black, short-dash black (dark-gray shaded region), dotted (light-gray shaded region) and gray-dashed contours, respectively. To the right of the black vertical line, axion DM is underabundant $\Omega h^2_a < 0.12$. In the left region $\Omega h^2_a = 0.12$, is possible for the pre-inflationary case depending on the level of tuning of the initial misalignment [see Eq.~\eqref{eq:relica}].}
    \label{fig:gagg}
\end{figure}
In Fig.~\ref{fig:gagg} we show $|g_{a\gamma\gamma}|$ as a function of the axion mass $m_a$ (bottom axis) and decay constant $f_a$ (top axis) [see Eq.~\eqref{eq:axionmass}]. The yellow band denotes the standard QCD-axion range $|E/N-1.92|\in[0.07,7]$~\cite{DiLuzio:2016sbl}. In the left plot, the benchmark KSVZ (hadronic)~\cite{Kim:1979if,Shifman:1979if} and DFSZ-I/II~\cite{Zhitnitsky:1980tq,Dine:1981rt} predictions are indicated by the solid red, blue, and green curves, respectively. Furthermore, the purple (orange) lines delimit the maximum and minimum $|g_{a\gamma\gamma}|$ obtained in case A/B (C), for the mediator representations listed in Table~\ref{tab:EN}. For case A/B [C], the largest coupling corresponds to $E/N=6$ [$E/N=16$], realized by $\Psi\sim(\mathbf{1},\mathbf{3},0)$ and $F\sim(\mathbf{3},\mathbf{3},-1/3)$ [$F\sim(\mathbf{3},\mathbf{3},-4/3)$] (see Tables~\ref{tab:mediators} and~\ref{tab:SU3Ncolor}). The smallest coupling occurs for $E/N=184/99$ [$E/N=64/33$], realized by $\Psi\sim(\mathbf{\bar{6}},\mathbf{3},0)$ and $F\sim(\mathbf{8},\mathbf{3},-1/3)$ [$\Psi\sim(\mathbf{8},\mathbf{1},0)$ and $F\sim(\mathbf{\bar{6}},\mathbf{1},-4/3)$]. In the right plot we show the $|g_{a \gamma \gamma}|$ predictions for the cosmologically conservative models \textbf{M}$_{1,2,3}$ identified in Sec.~\ref{sec:axiondarkmatter}, which feature $N_{\text{DW}}=1$ and a lightest stable particle that is either a fermionic or scalar SM singlet. The predictions for \textbf{M}$_{1}$ and \textbf{M}$_{2}$ are the same. Namely, for case A/B (C) we have $E/N= 2/3$ ($E/N= 32/3$) indicated by the green (magenta) oblique line. Furthermore, for \textbf{M}$_{3}$, case A/B (C) leads to $E/N=14/3$ ($E/N=8/3$), shown by the cyan (blue) line. Notice that the predictions of \textbf{M}$_{1,2}^{\text{A/B}}$ and \textbf{M}$_{3}^{\text{C}}$ match the ones of the paradigmatic DFSZ II and I models, respectively.

In the figure we show the current bounds and projected sensitivities from helioscopes and haloscopes. The CAST~\cite{CAST:2017uph} helioscope experiment excludes the burgundy-shaded region, while IAXO is expected to reach $g_{a\gamma\gamma}\sim(10^{-12}\!-\!10^{-11})\,\mathrm{GeV}^{-1}$ and probe the benchmark QCD-axion models (red, green and blue lines) around $m_a\sim0.1~\mathrm{eV}$ (burgundy dashed contour). Haloscopes target non-relativistic axions, assuming they constitute DM. As discussed in Sec.~\ref{sec:axiondarkmatter}, the misalignment mechanism yields the observed DM relic abundance only to the left of the black vertical line depending on the level of tuning of the initial misalignment angle [see Eq.~\eqref{eq:relica}], while to the right, axions are underabundant. Note that, the preferred cosmological scenarios \textbf{M}$_{1,2,3}$, can be more predictive since post-inflationary axion DM, free of DWs, becomes viable. Namely, given current numerical simulations to obtain $\Omega_a h^2 = \Omega_{\text{DM}} h^2$, requires $f_a \in [5 \times 10^9 , 3 \times 10^{11}]$~GeV~\cite{Kawasaki:2014sqa,Klaer:2017ond,Gorghetto:2020qws,Buschmann:2021sdq,Benabou:2024msj}. As mentioned before in Sec.~\ref{sec:axiondarkmatter}, beyond \textbf{M}$_{1,2,3}$ the remaining models, feature $N_{\text{DW}} >1$, as well as stable colored and/or electrically charged relics. Thus for these models it is judicious to assume pre-inflationary axion DM, where inflation washes out topological defects and dangerous stable relics. Nonetheless, post-inflationary axion DM can still be viable, ableit with a few caveats. This possibility may require to introduce small bias terms in the scalar potential that render the DW-string network unstable~\cite{Sikivie:1982qv}, and a mechanism to suppresses the abundance of dangerous stable relics or allow them to decay into ordinary matter~\cite{Nardi:1990ku,DiLuzio:2016sbl,DiLuzio:2017pfr}. In the figure, the haloscopes, ADMX~\cite{ADMX:2018gho,ADMX:2019uok,ADMX:2021nhd}, RBF~\cite{DePanfilis:1987dk}, CAPP~\cite{CAPP:2020utb}, and HAYSTAC~\cite{HAYSTAC:2020kwv} exclude the black-shaded region. Among these, ADMX has already reached the KSVZ (red line) and DFSZ (blue/green line) predictions near $m_a\sim3~\mu\mathrm{eV}$, and the forthcoming ADMX generations (black dashed contour) aim for sensitivities down to $g_{a\gamma\gamma}\sim10^{-16}\,\mathrm{GeV}^{-1}$ over $1~\mu\mathrm{eV}\lesssim m_a\lesssim100~\mu\mathrm{eV}$. In addition, ALPHA~\cite{Lawson:2019brd,Wooten:2022vpj,ALPHA:2022rxj} (light-gray band) and MADMAX~\cite{Beurthey:2020yuq} (dark-gray band) are projected to cover $20~\mu\mathrm{eV}\lesssim m_a\lesssim120~\mu\mathrm{eV}$ and $50~\mu\mathrm{eV}\lesssim m_a\lesssim120~\mu\mathrm{eV}$, respectively. Finally, ABRACADABRA~\cite{Ouellet:2018beu} (gray dashed contour) is expected to probe $m_a\lesssim1~\mu\mathrm{eV}$ and reach the yellow shaded QCD-axion band.

\newpage

%%%%%%%%%%%%%%%%%%%%%%%%%%%%%%%%%%%%%%%%%%%%%%%%%%%%%%%%%%%%%%%%%%%%%%%%%%%%%
\section{Concluding remarks}
\label{sec:concl}
%%%%%%%%%%%%%%%%%%%%%%%%%%%%%%%%%%%%%%%%%%%%%%%%%%%%%%%%%%%%%%%%%%%%%%%%%%%%%

In this work, we propose a framework in which the PQ symmetry is identified with $B+L$ symmetry, while proton decay is generated at one-loop level through the exchange of vector-like fermions $\Psi$, $F$ and scalar mediators $S$, $\chi$, as shown in Figs.~\ref{fig:Diagram_Operator} and~\ref{fig:Diagram_Proton_Decay}. The PQ anomaly sector, composed of the vector-like fermions $\Psi$ and $F$, chirally charged under U(1)$_{\text{PQ}}$, solves the strong CP problem within a KSVZ-type axion framework. After spontaneous PQ-symmetry breaking, a residual $\mathcal{Z}_2$ symmetry remains unbroken, under which the mediator fields $\Psi$, $F$, $S$, and $\chi$ are odd, guaranteeing that proton decay is forbidden at tree level.

We study UV completions of the operator $u_R u_R d_R e_R \sigma^{\ast 2}$, which, after PQ-symmetry breaking, when $\sigma$ acquires a non-zero VEV, induces proton decay at one loop. Different realizations, characterized by distinct vector-like fermion and scalar representations under the SM gauge group, lead to different proton and neutron decay channels shown in Fig.~\ref{fig:Diagram_Proton_Decay} and summarized in Table~\ref{tab:boundsProtonDecay}. In addition to the standard channel $p \rightarrow e^+ \pi^0$, the framework predicts axion-emitting modes such as $p \rightarrow e^+ \pi^0 a$, though these are strongly suppressed by the PQ-breaking scale.

Different representation assignments of the mediator fields $\Psi,F,S$ and $\chi$ under the SM gauge group lead to distinct predictions for the axion-to-photon coupling. This allows to experimentally distinguish the different realizations of our framework at haloscope and helioscope experiments such as IAXO, ADMX, and MADMAX, as shown in Fig.~\ref{fig:gagg}. 

We also discuss axion DM in pre- and post-inflationary cosmological scenarios. Due to the residual $\mathcal{Z}_2$ symmetry, the lightest odd field among $\Psi,F,S$ and $\chi$ can yield long-lived colored and/or electrically charged relics. Furthermore, most realizations have a DW number $N_{\text{DW}} >1$, which would produce axion string and DW networks, motivating a pre-inflationary scenario where such defects are inflated away. In this case, axions can account for the observed cold DM abundance for $\theta_0 \sim \mathcal{O}(1)$ and $f_a \sim 5 \times 10^{11}$ GeV, via the misalignment mechanism.

In summary, our framework promotes the accidental global $B+L$ symmetry already present in the SM to a PQ symmetry, providing a UV completion of the KSVZ axion in which proton decay arises radiatively, thereby connecting the stability of the proton with the axion solution to the strong CP problem.

%%%%%%%%%%%%%%%%%%%%%%%%%%%%%%%%%%%%%%%%%%%%%%%%%%%%%%%%%%%%%%%%%%%%%%%%%%%%%
\begin{acknowledgments}
HBC would like to thank F.R. Joaquim and R. Srivastava for carefully reading the manuscript and for helpful suggestions, as well as J. R. Rocha for valuable discussions. HBC thanks CFTP (Lisbon) for hospitality and financial support during the final stage of this work. 
\end{acknowledgments}
%%%%%%%%%%%%%%%%%%%%%%%%%%%%%%%%%%%%%%%%%%%%%%%%%%%%%%%%%%%%%%%%%%%%%%%%%%%%%

%%%%%%%%%%%%%%%%%%%%%%%%%%%%%%%%%%%%%%%%%%%%%%%%%%%%%%%%%%%%%%%%%%%%%%%%%%%%%

%%%%%%%%%%%%%%%%%%%%%%%%%%%%%%%%%%%%%%%%%%%%%%%%%%%%%%%%%%%%%%%%%%%%%%%%%%%%%

\end{document}